\documentclass[12pt]{article}
\newcommand{\bkt}[1]{\left\langle#1\right\rangle}

\newcommand{\gs}{\Phi_{0}}
\newcommand{\ret}{\nonumber\\}
\newcommand{\ep}{|\varepsilon|}
\newcommand{\EGS}{E_{\rm GS}}

\begin{document}
\noindent
{\large\bf Low-lying excitations in one-dimensional lattice electron systems}

\bigskip\noindent
Hal Tasaki\footnote{
Department of Physics, Gakushuin University,
Mejiro, Toshima-ku, Tokyo 171-8588, Japan
}

\begin{abstract}
We consider a general one-dimensional tight-binding electron model which has a period $P$.
For any filling factor $\nu$ such that $P\nu$ is non-integral, we prove that the model in the infinite volume limit has either a symmetry breaking or a unique ground state with gapless excitations.
The proof is based on the idea of Yamanaka, Oshikawa and Affleck, who extended the Lieb-Schultz-Mattis argument to electron systems.
\end{abstract}

In spite of considerable interest, rigorous results for strongly interacting electron systems are still rare \cite{Tasaki}.
Some years ago, Yamanaka, Oshikawa, and Affleck \cite{YOA} proposed an  interesting and  fully nonperturbative theorem about low-lying excitations in one-dimensional lattice electron systems.
They showed that the celebrated  Lieb-Schultz-Mattis theorem \cite{LSM} originally stated for one-dimensional quantum spin systems may be extended to electron systems.

In \cite{YOA}, however, only the finite volume version of the theorem was explicitly proved.
They constructed a low-lying excited state in which the excitation extends over the whole periodic lattice.
As was pointed out by Affleck and Lieb \cite{AL}, such a trial state may not correspond to a physical excited state since there is a possibility that the state converges to the ground state in the infinite volume limit.
In order to fully characterize the low energy physics of the model, one needs a trial state with a local excitation, i.e., an excited state which differs from the ground state only in a finite region of the lattice \cite{phys}.
It was shown in \cite{AL} that, with such a localized excited state, one can proceed to prove a rigorous theorem about the ground state and the low-lying excitations in the infinite volume limit.

As for quantum spin systems, Affleck and Lieb \cite{AL} constructed excited states with local excitations, and proved that they are orthogonal to the ground state. 
But the authors of \cite{YOA} have noticed that the same proof of orthogonality fails in the electron systems since the latter have lesser symmetry.

The purpose of the present paper is to fill this gap.
We shall develop a new method which makes use of the charge correlation function and construct excited states with local excitations.
The method is fairly general and can be applied to many situations where the simpler method of \cite{AL} does not apply.
Our result was briefly mentioned in \cite{YOA}.

\paragraph*{Model and main result:}
Let us state and prove the result in the most general form.

We fix a positive integer $P$, which is the period of the system.
For an integer $N$ (which will finally become infinite), consider the chain $\Lambda=\{0,1,\ldots,NP-1\}$  with a periodic boundary condition.
We define as usual the annihilation operator $c_{j,\sigma}$ for the electron at $j\in\Lambda$ with spin $\sigma=\uparrow$, $\downarrow$.
We consider a general tight-binding electron system with the Hamiltonian
\begin{equation}
H=-\sum_{j,j'\in\Lambda}\sum_{\sigma=\uparrow,\downarrow}
t_{j,j'}\,
c^\dagger_{j,\sigma}\,c_{j',\sigma}
+\sum_{j,j'\in\Lambda}V_{j,j'}\,\hat{n}_j\hat{n}_{j'},
\label{e:H}
\end{equation}
where $\hat{n}_j=\sum_{\sigma}c^\dagger_{j,\sigma}c_{j,\sigma}$ is the number operator at site $j$.
The hopping amplitudes are real and satisfy $t_{j,j'}=t_{j',j}$.
Note that the diagonal component $t_{j,j}$ represents the single-body potential.
We assume that both the hopping and the interaction have period $P$, i.e., $t_{j+P,j'+P}=t_{j,j'}$ and $V_{j+P,j'+P}=V_{j,j'}$ .
We also assume that the hopping is short-ranged; there is a constant $R$ (which is independent of the system size $N$) and one has $t_{j,j'}=0$ whenever $|j-j'|>R$.
The potential may be long ranged.

Fix the filling factor $\nu$ such that $0<\nu<1$.
We consider the Hilbert space with $2M$ electrons where $M$ is the integer closest to $\nu N P$.
We assume that the ground state $\Phi_0$  is unique for each finite $N$, and consider the infinite volume limit $N\to\infty$  with the fixed $\nu$.

\bigskip
\noindent
{\em Theorem:}\/
When $P\nu$ is not an integer,  one of the following two alternatives is valid:
{\em i)~there is a symmetry breaking, and the infinite volume ground states are not unique;
 ii)~there are gapless excitations above the unique infinite volume ground state}\/.
 
 \bigskip
In other words, we can rigorously rule out the third possibility that {\em
iii)~the infinite volume ground state is unique, and there is a finite gap above it.
}

In a noninteracting electron system, it is an easy exercise to show that, for non-integral  $P\nu$, one never has iii) and indeed has i).
This theorem shows that the situation remains essentially the same no matter how strong the interactions and correlations are.
Let us again stress the perfect rigor and the  fully non-perturbative nature of the result.

\paragraph*{Proof:}
Our goal is to show the following.
For an arbitrary (small) $\Delta E>0$, there exist finite $L$ and $N_0$ such that for any $N\ge N_0$ we can find a trial state $\Psi'$ with the following properties.
The state $\Psi'$ is orthogonal to the ground state $\gs$, has energy not larger than $\EGS+\Delta E$ (where $\EGS$ is the ground state energy), and differs from $\gs$ only in the interval $\{0,1,\ldots,LP-1\}$.

Consider the finite chain, and denote by $\bkt{\cdots}=\bkt{\gs,(\cdots)\gs}$ the ground state expectation.
Take $L$ such that $L<N$, and
define the density operator
\begin{equation}
\hat{\rho}_L=(LP)^{-1}\sum_{j=0}^{LP-1}\hat{n}_{j,\uparrow},
\label{e:rho}
\end{equation}
for the up-spin electrons, where $\hat{n}_{j,\uparrow}=c^\dagger_{j,\uparrow}c_{j,\uparrow}$ is the number operator.
Consider the density fluctuation 
\begin{eqnarray}
f(L)&=&\bkt{(\hat{\rho}_L-\nu)^{2}}
\nonumber\\
&=&\frac{1}{(LP)^{2}}
\sum_{j,j'=0}^{LP-1}
\bkt{\hat{n}_{j,\uparrow}\hat{n}_{j',\uparrow}}-\bkt{\hat{n}_{j,\uparrow}}\bkt{\hat{n}_{j',\uparrow}}.
\label{e:fL}
\end{eqnarray}
We have noted that there are $M$ up-spin-electrons in the unique ground state, which is necessarily a spin-singlet.
Let us assume here that $f(L)\to0$ as $L\to\infty$.
Otherwise there is a long range charge order, and we are done with the situation i) above.

Define the unitary ``twist'' operator by
\begin{equation}
U=\exp\{2\pi i\sum_{j=0}^{LP-1}([j/P]+1)(\hat{n}_{j,\uparrow}/L)\},
\label{e:U}
\end{equation}
where $[\cdots]$ is the Gauss symbol.
We define our trial state as $\Psi=U\gs$.
Let $T$ be the translation by $P$.
Then an explicit calculation shows that
$T\Psi=\exp\{-2\pi iP\hat{\rho}_L\}\Psi$.
Since $T\gs=\gs$, we have
\begin{eqnarray}
2\bkt{\gs,\Psi}&=&\bkt{(T^{-1}+1)\gs,\Psi}
\nonumber\\
&=&\bkt{\gs,(T+1)\Psi}
\nonumber\\
&=&\bkt{\gs,(e^{-2\pi iP\hat{\rho}_L}+1)\Psi}
\nonumber\\
&=&\bkt{(e^{2\pi iP\hat{\rho}_L}+1)\gs,\Psi}.
\label{e:gP}
\end{eqnarray}
Then from the Schwarz inequality we get
\begin{eqnarray}
    |\bkt{\gs,\Psi}|^{2}
	 & \le & 
	 \bkt{\frac{e^{2\pi iP\hat{\rho}_L}+1}{2}\gs,
	 \frac{e^{2\pi iP\hat{\rho}_L}+1}{2}\gs}
	 \bkt{\Psi,\Psi}
	\ret
	 & = & \bkt{\gs,\frac{\cos(2\pi P\hat{\rho}_L)+1}{2}\gs}.
	\label{main}
\end{eqnarray}

Now for each $\nu$ such that $P\nu$ is not an integer, we can choose constants 
$0<\alpha<1$ and $\beta>0$ such that
$\{\cos(2\pi Px)+1\}/2\le\alpha+\beta(x-\nu)^{2}$
holds for any real $x$.
This is indeed an elementary fact best proved by drawing a graph.
Then (\ref{main}) implies
\begin{eqnarray}
	|\bkt{\gs,\Psi}|^{2} 
	& \le & 
	 \bkt{\alpha+\beta(\hat{\rho}_L-\nu)^{2}}
	\ret
	&\le&
	 \alpha+\beta f(L)
\le
	 1-\delta,
	\label{overlap}
\end{eqnarray}
where $\delta>0$ is a constant.
The final bound is valid for sufficiently large $L$.
Unlike in \cite{AL} we are not able to show that $\Psi$ and $\gs$ are orthogonal.
But the above weaker estimate is sufficient for our purpose.

Decompose the trial state $\Psi$ as
\begin{equation}
\Psi=\sqrt{1-\ep^{2}}\Psi'+\varepsilon\gs,
\label{e:dec}
\end{equation}
with $\varepsilon=\bkt{\gs,\Psi}$, $\bkt{\Psi',\Psi'}=1$, and
$\bkt{\Psi',\gs}=0$.
The resulting $\Psi'$ is our new trial state.
We have
\begin{equation}
\bkt{\Psi,H\Psi}=(1-\ep^{2})\bkt{\Psi',H\Psi'}+\ep^{2}\EGS,
\end{equation}
where $\EGS$ is the ground state energy.
On the other hand, the standard estimate as in \cite{LSM,YOA,AL} shows
\begin{equation}
\bkt{\Psi,H\Psi}\le\EGS+(\gamma/L),
\end{equation}
with some constant $\gamma$.
By combining these two, and using $\ep^{2}\le 1-\delta$ (which is 
(\ref{overlap})), we get
\begin{eqnarray}
	\bkt{\Psi',H\Psi'}& \le &
	\frac{(1-\ep^{2})\EGS+\gamma L^{-1}}{1-\ep^{2}}
	\ret
	 &= &\EGS+\frac{\gamma}{1-\ep^{2}}\frac{1}{L}
	 \ret&\le&
	 \EGS+\frac{\gamma}{\delta}\frac{1}{L},
\end{eqnarray}
which shows the existence of the desired low lying excitation with a 
large but finite support of length $L$.

Now, by following the argument in \cite{AL}, we see that only i) or ii) is possible.
Note that, although we have already separated the case with a long range charge order, there still is a possibility that the system develops other types of ordering.
Unfortunately the present argument is not strong enough to specify the type of ordering or to distinguish between the cases i) and ii).

I thank  Ian Affleck , Tohru Koma, Masaki Oshikawa, and Masanori Yamanaka for useful discussions.


\begin{thebibliography}{10}

\bibitem{Tasaki}
H. Tasaki, J. Phys.: Cond. Mat. {\bf 10} 4353 (1998), cond-mat/9512169.

\bibitem{YOA}
M. Yamanaka, M. Oshikawa, and I. Affleck, Phys. Rev. Lett. {\bf 79}, 1110 (1997).

\bibitem{LSM}
E.~H. Lieb, T. Schultz, and D.~J. Mattis, Ann. Phys. (N.Y.) {\bf 16},  407
  (1961).


\bibitem{AL}
I. Affleck and E.~H. Lieb, Lett. Math. Phys. {\bf 12},  57  (1986).

\bibitem{phys}
Note that this requirement for the locality of excitation is physically natural.
In experiments, one is rarely interested in very special excitations which extends over the whole sample.
Observable low-lying excitations are in general localized in the sample.


\end{thebibliography}
\end{document}